\documentclass[aps,prl,twocolumn,showpacs]{revtex4}

\usepackage{amssymb,graphicx,times}

\begin{document}

\title{Polynomial growth in branching processes\\ with diverging
reproductive number}

\author{Alexei Vazquez}

\affiliation{Department of Physics and Center for Complex Network
Research, University of Notre Dame, IN 46556, USA}

\date{\today}

\begin{abstract}

We study the spreading dynamics on graphs with a power law degree
distribution $p_k\sim k^{-\gamma}$ with $2<\gamma<3$, as an example of a
branching process with diverging reproductive number. We provide evidence
that the divergence of the second moment of the degree distribution
carries as a consequence a qualitative change in the growth pattern,
deviating from the standard exponential growth. First, the population
growth is extensive, meaning that the average number of vertices reached
by the spreading process becomes of the order of the graph size in a time
scale that vanishes in the large graph size limit. Second, the temporal
evolution is governed by a polynomial growth, with a degree determined by
the characteristic distance between vertices in the graph. These results
open a path to further investigation on the dynamics on networks.

\end{abstract}


\maketitle

\bibliographystyle{apsrev}

Branching processes model the evolution of populations whose elements
reproduce generating new elements \cite{harris02,kimmel02}, such as a
population of physical particles \cite{bharucha54,mekjian01}, cells
\cite{kimmel02}, or infected individuals \cite{pv00}. A key magnitude
determining the dynamical evolution of the population size is the average
reproductive number $\tilde{R}$, giving the number of secondary particles
generated by a primary particle. When $\tilde{R}<1$ the average number of
new elements decreases exponentially, while it grows exponentially when
$1<\tilde{R}<\infty$ \cite{harris02}. On the other hand, it has been
recently found that $\tilde{R}$ may be unbounded for branching
processes taking place on graphs with a power law degree distribution
\cite{pv00,pv01,nsw01,n02c,ceah00}, where by unbounded we mean that
$\tilde{R}$ diverges with increasing graph size. This observation is
extremely important since several graphs representing interactions among
human or computers are characterized by a power law degree distribution
\cite{bajb00a,pvv01,liljeros01,eckmann04,emb02}, requiring us to consider
branching processes with an unbounded average reproductive number.

Barth{\'e}lemy {\it et al} \cite{bbv04} have recently studied the
spreading dynamics of an infectious disease on a graph with a power law
degree distribution. Using a mean-field approach they obtained that the
average number of infected vertices grows exponentially in time with a
characteristic time $\tau\sim \langle k^2\rangle^{-1}$, where $\langle
k^2\rangle$ is the second moment of the degree distribution $p_k$. For
graphs where $p_k\sim k^{-\gamma}$ with $2<\gamma<3$ the second moment
diverges and $\tau\rightarrow0$ with increasing graph size, predicting
that all vertices will be instantaneously infected \cite{bbv04}. A disease
that spreads at constant rate, however, cannot spread to all vertices in a
time scale much smaller than the inverse of the spreading rate, indicating
that the predicted exponential growth should not dominate the system's 
dynamics.

In this work we study branching processes with an unbounded average
reproductive number using a spreading process on a graph as a case
study. When the degree distribution has the power law tail $p_k\sim
k^{-\gamma}$ with $2<\gamma<3$ we obtain that the exponential regime is
followed by a polynomial growth in time, a result that is completely
unexpected based on previous mathematical studies. We also show that both
the characteristic time separating the exponential and polynomial regimes
and the polynomial degree depend on the characteristic distance between
vertices.  More important, in the limit of infinite graph sizes the
exponential regime is virtually absent, indicating that the polynomial
regime is a novel and characteristic feature of the spreading dynamics on
graphs with degree exponent $2<\gamma<3$, and more generally of branching
processes with an unbounded average reproductive number.

Consider a spreading process on a graph with a tree-like structure.  At
$t=0$ a vertex selected at random is infected by a ``virus'', which can
then propagate to other vertices through the graph edges. The causal tree
representing the spreading process can be modeled as a branching process.  
Each vertex in the causal tree represents an infected vertex in the
original graph, and each arc in the causal tree represents the generation
of a secondary infected vertex from a primary infected vertex. The
out-degree of a vertex in the causal tree gives the number of other
vertices it infects, {\it i.e.} its reproductive number. In turn, the
length of an arc A$\rightarrow$B in the causal tree gives the generation
time, the time elapsed from the infection of the primary case A to the
infection of the secondary case B. Finally, the vertex generation coincides
with the topological distance from the first infected vertex, the root, in
the original graph.

We assume that the reproductive numbers are independent random variables
with the probability distribution $q_k^{(d)}$ and average reproductive
number $R^{(d)}=\sum_k q_k^{(d)}k$, parametrized by the generation $d$.
The parametrization by $d$ is introduced to take into account that the
degree distribution may change significantly from generation to generation
\cite{nsw01,cohen03b}.  We also assume that the generation times are
independent random variables with the distribution $G^{(d)}(\tau)$ and the
probability density $g^{(d)}(\tau)=dG^{(d)}(\tau)/d\tau$. Let
$P_N^{(d)}(t)$ be the probability distribution of the number of vertices
$N$ that are found at time $t$ in a branch of the causal tree, given that
branch is rooted at a vertex at generation $d$. Because of the tree
structure we can write the recursive relation

\begin{widetext}
\begin{equation}
\displaystyle
P^{(d)}_N(t) = 
\sum_{k=0}^\infty q^{(d)}_k
\sum_{N_1=0}^\infty\dots\sum_{N_k=0}^\infty
\delta_{\sum_{i=1}^kN_i +1, N}
\prod_{i=1}^k \left[ \int_0^t dG^{(d)}(\tau) P^{(d+1)}_{N_i}(t-\tau)
+ \delta_{N_i,0}\left( 1 - G^{(d)}(t)\right) \right]\ ,
\label{Pd}
\end{equation}
\end{widetext}

\noindent with the boundary condition $P^{(D)}_N(t)=\delta_{N,1}$, where
$D$ is the maximum distance between two vertices on the graph. The sum
over $k$ runs over the possible reproductive numbers of the reference
vertex, while the sum over $N_i$, $i=1,\ldots,k$, runs over the possible
number of infected vertices in the branch rooted at the $i$th neighbor of
the reference vertex. These sums are then restricted by the Kronecker
delta to configurations satisfying $1+\sum_{i=1}^kN_i=N$.  Finally, within
the $[\cdots]$ we have the probability that the branch rooted at the
$i$-th neighbor has $N_i$ infected vertices at time $t-\tau$, averaged
over the generation time distribution $G(\tau)$. The product structure in
(\ref{Pd}) suggests the use of the generating functions

\begin{equation}
H^{(d)}(x) = \sum_{k=0}^\infty q^{(d)}_k x^k\ ,
\label{G0}
\end{equation}

\begin{equation}
F^{(d)}(x,t) = \sum_{N=0}^\infty P^{(d)}_N(t) x^N \ ,
\label{F0d}   
\end{equation}

\noindent for the reproductive number and the number of infected vertices,
respectively. From (\ref{Pd})-(\ref{F0d}) we obtain

\begin{widetext}
\begin{equation}
\displaystyle
F^{(d)}(x,t) = x H^{(d)} \left(\int_0^t dG^{(d)}(\tau) F^{(d+1)}(x,t-\tau)
+ 1 - G^{(d)}(t)\right)\ ,
\label{Fd}
\end{equation}
\end{widetext}

\noindent with the boundary condition $F^{(D)}(x)=x$. From this equation
we obtain the average number of infected vertices on the branch rooted at
a vertex at generation $d$,

\begin{equation}
N^{(d)}(t) =
\frac{\partial F^{(d)}(1,t)}{\partial x}\ ,
\label{NF}
\end{equation}

\noindent with the boundary condition $N^{(D)}(t)=1$. Iterating this
equation from $d=D$ to $d=0$ we obtain the average number of infected
vertices at time $t$, $N^{(0)}(t)$, and the average number of new vertices
infected between $t$ and $t+dt$, $n(t)=dN^{(0)}(t)/dt$, resulting in

\begin{equation}
n(t) = \sum_{d=1}^D z_d
\left( g^{(0)}\star g^{(1)}\star\cdots\star g^{(d)}(t) \right)
\ ,
\label{Nt}
\end{equation}   

\noindent where 

\begin{equation}
z_d = \prod_{l=0}^d R_l
\label{zd}
\end{equation}

\noindent is the average number of vertices at generation $d$, and the
second factor is the probability that the infection has reached a vertex
at generation $d$, where $\star$ denotes the convolution operation, for
instance $g^{(0)}\star g^{(1)}(t)=\int_0^t d\tau
g^{(0)}(\tau)g^{(1)}(t-\tau)$.

Next we consider the cases when: ({\it i}) the reproductive number of
vertices other than the root has the same statistical properties, {\it
i.e.} $R^{(0)}=R$ and $R^{(d)}=\tilde{R}$ for $d\geq1$, and ({\it ii}) the
infection is transmitted from an infected vertex to a susceptible (not yet
infected) vertex at constant rate $1/T_{\rm G}$. This last assumption
corresponds to an exponential distribution of generation times
$G^{(d)}(\tau)=1-\exp(-t/T_{\rm G})$, with average generation time $T_{\rm
G}$. Under these approximations from (\ref{Nt}) and (\ref{zd}) we obtain

\begin{equation}
n(t) = \frac{R}{T_{\rm G}}
\exp\left(-\frac{t}{T_{\rm G}}\right)
\sum_{d=1}^D \frac{1}{(d-1)!}
\left(\frac{\tilde{R}t}{T_{\rm G}}\right)^{d-1}\ ,
\label{nt1}
\end{equation}

\noindent The sum in (\ref{nt1}) is the Taylor series expansion of
$\exp(\tilde{R}t/T_{\rm G})$, up to the $D-1$ order.  It actually
approximates an exponential function depending on the ratio of $t/\tau_0$,
where

\begin{equation}
\tau_0 = T_{\rm G}\frac{D}{\tilde{R}}\ .
\label{tau0}
\end{equation}

\noindent When $t\ll\tau_0$ we obtain

\begin{equation}
n(t) \approx
\frac{R}{T_{\rm G}}
\exp\left( (\tilde{R}-1) \frac{t}{T_G} \right)\ ,
\label{Nttheta1}
\end{equation}

\noindent becoming an exponential growth for $\tilde{R}>1$
\cite{harris02,bbv04}. In contrast, when $t\gg\tau_0$ we obtain
a polynomial growth followed by an exponential decay:

\begin{equation}
n(t) \approx 
\frac{R\tilde{R}^{D-1}}{ T_{\rm G} (D-1)!}
\left( \frac{t}{ T_{\rm G} } \right)^{D-1}
\exp\left( -\frac{t}{T_{\rm G}} \right)\ .
\label{Nttheta2}
\end{equation}

In general the time scale $\tau_0$ depends on the graph size $N_0$. For
random graphs with an arbitrary degree distribution $q^{(0)}_k=p_k$ and
$q^{(d)}_k=(k-1)p_{k-1}/\langle k\rangle$ for $d>0$ \cite{nsw01},
resulting in $R\sim\langle k\rangle$ and $\tilde{R}\sim \langle
k^2\rangle$, where $\langle k\rangle$ and $\langle k^2\rangle$ are the
first and second moments of the degree distribution. In this case we 
obtain the following scenarios:

({\it i}) When the tail of the degree distribution decays faster than
$p_k\sim k^{-3}$ the diameter scales as $D\sim\log N_0$ \cite{nsw01},
while $\tilde{R}$ is constant or approaches a constant in the large graph
size limit. Thus, from (\ref{tau0}) it follows that

\begin{equation}
\tau_0 \sim \frac{T_{\rm G}}{\tilde{R}}\log N_0\ .
\label{tau0i}
\end{equation}

\noindent In this case the exponential growth last till $t\sim\tau_0$,
where $\tau_0\rightarrow\infty$ when $N_0\rightarrow\infty$.

({\it ii}) When the degree distribution has the power law tail $p_k\sim
k^{-\gamma}$ with $2<\gamma<3$, the diameter $D$ increases at most as
$\log N_0$ \cite{chung02,bollobas03,cohen03a}, while $\tilde{R}\sim
N_0^{(3-\gamma)/(\gamma-1)}$. Thus, from (\ref{tau0}) it follows that

\begin{equation}
\tau_0 \sim T_{\rm G} \frac{\log N_0}{N_0^{(3-\gamma)/(\gamma-1)}}\ .
\label{tau0ii}
\end{equation}

\noindent 

\noindent The initial exponential growth is thus a finite size effect
restricted to $t\ll\tau_0$, where $\tau_0\rightarrow0$ when
$N_0\rightarrow\infty$. Following this vanishing time window the number of
infected vertices is already of the order of the graph size $N_0$ 
($R\tilde{R}\sim N_0$) and its
temporal evolution is polynomial (\ref{Nttheta2}), with a degree
determined by the characteristic distance between vertices in the
underlying graph.

\begin{figure}

\centerline{\includegraphics[width=3in]{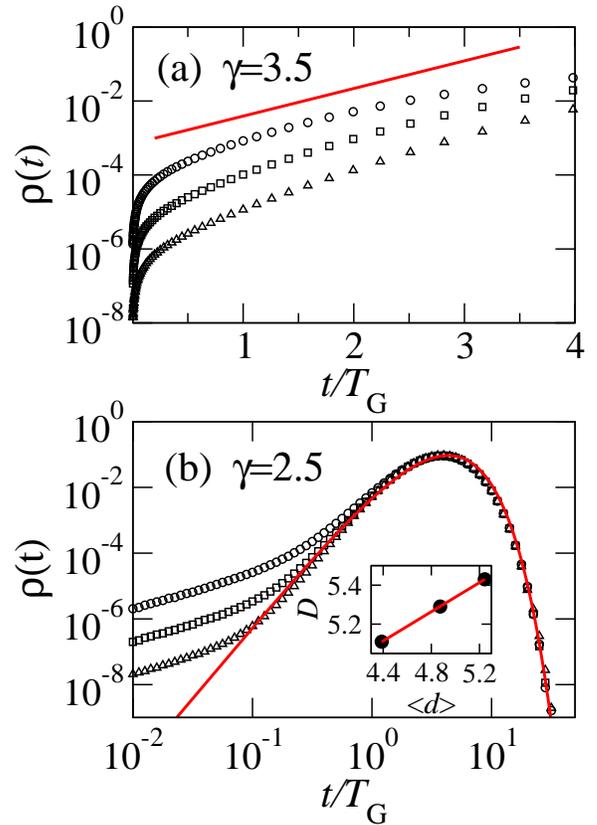}}

\caption{Fraction of infected nodes $\rho(t)=n(t)/N_0$ as a function of
time resulting from SI model simulations on random graphs with a power law
degree distribution $p_k=Ak^{-\gamma}$, with $\gamma=3.5$ (a) and
$\gamma=2.5$ (b). Different symbols correspond with different graph sizes:
$N_0=1000$ (circles), 10,000 (squares) and 100,000 (triangles). (a) For
$\gamma=3.5$ the spreading dynamics is characterized by an exponential
growth (line), as predicted by (\ref{Nttheta1}). (b) For $\gamma=2.5$ the
number of new infections is better described by (\ref{Nttheta2}) (line).
There are some deviations at short times, but they get reduced with
increasing the graph size. The inset shows the exponent $D$ resulting from
the fit of (\ref{Nttheta2}) as a function of the average distance $\langle
d\rangle$ between two nodes in the graph. The increase in $\langle
d\rangle$ is obtained by increasing the network size from $N_0=1000$ to
10,000, and 100,000. The line emphasizes the linear scaling between $D$
and $\langle d\rangle$.}

\label{fig1}
\end{figure}

To check the validity of our calculations we perform numerical simulations
of the susceptible infected (SI) model on random graphs with a power law
degree distribution $p_k=Ak^{-\gamma}$. Within this model, vertices can be
in two states, susceptible or infected, and infected vertices transmit the
infection to each of its neighbors at a constant rate $1/T_{\rm G}$
\cite{anderson91}. We generate random graphs with a power law degree
distribution using the algorithm proposed in \cite{park03}. Then we
generated single outbreaks on these graphs starting from one infected
vertex. Finally, we take averages over 10,000 outbreaks starting from
randomly selected vertices, and over 100 graph realizations. 

\begin{figure}

\centerline{\includegraphics[width=3in]{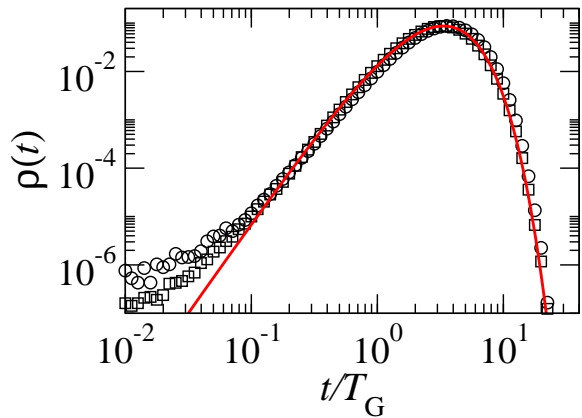}}

\caption{ Fraction of infected nodes $\rho(t)=n(t)/N_0$ as a function of
time resulting from SI model simulations on AS networks, of September 1997
with $N_0=3015$ (circles) and of October 2001 with $N_0=10515$ (squares).
The line is a fit to (\ref{Nttheta2}) resulting in $D=4.7\pm0.1$.}

\label{fig2}
\end{figure}

When $\gamma>3$ the spreading dynamics is better described by an initial
exponential growth (Fig. \ref{fig1}a), in agreement with (\ref{Nttheta1})  
and previous mathematical approaches \cite{may88,anderson91,bbv04}. In
contrast, when $2<\gamma<3$ the spreading dynamics is better described by
(\ref{Nttheta2}) (Fig. \ref{fig1}b), and the exponent $D$ resulting from
the fit to the numerical data scales linearly with the average distance
between nodes (see inset of Fig. \ref{fig1}b). In a more realistic
scenario, we use the SI model to simulate the spreading of a routing table
error on the Autonomous System (AS) network representation of the Internet
\cite{pv04}. This network is characterized by a power law degree
distribution with $\gamma\approx2.1$ \cite{pvv01}, but it also exhibits
degree-degree correlations \cite{pvv01} and a large degree dependent
clustering coefficient \cite{vpv02a}. Yet, the average number of new
infections is well fitted by (\ref{Nttheta2}), indicating that our
predictions are also valid for graphs that are not random as well (see
Fig. \ref{fig2}).

With relevance to the spreading of computer virus and worms among email
users, there is empirical evidence indicating that Email networks are
characterized by a power law degree distribution with $2<\gamma<3$
\cite{eckmann04,emb02}. The transmission rates of computer viruses are,
however, of the order of their typical detection times, making difficult
the empirical observation of the initial epidemic growth. With relevance
to sexually transmitted diseases, there are several reports indicating
that the network of sexual contacts is characterized by a power law degree
distribution \cite{liljeros01,jones03a,schneeberger04}, with an exponent
$\gamma>3$ for some communities and $2<\gamma<3$ for others.  This fact
together with the results obtained in this work represent a possible
explanation for the observation of both exponential and polynomial HIV
epidemic growth in different populations
\cite{may88,brookmeyer94,szendroi04}. The available data is, however, not
sufficient to make a definitive conclusion.

In a more general perspective, our results indicate that the degree
statistics is not sufficient to characterize the spreading dynamics, and
probably other dynamical processes, taking place on graphs with a power
law degree distribution with exponent $2<\gamma<3$. To determine the
characteristic time $\tau_0$ and the polynomial degree we need the
characteristic distance between vertices in the graph as well. The amount
of information needed to determine the distance between vertices is,
however, more difficult to collect, in principle requiring the complete
mapping of the graph. In this respect, the development of realistic graph
models that can accurately represent the real graphs will be extremely
valuable, allowing us to characterize the distance statistics from the
degree statistics.

\noindent {\bf Acknowledgments:} I thank A.-L. Barab\'asi, A. Vespignani,
and E. Almaas for helpful comments and suggestions. This work was
supported by NSF ITR 0426737, NSF ACT/SGER 0441089 awards.

%\bibliography{network}

\end{document}